\documentclass[preprint,aps,tightenlines,showkeys,nofootinbib]{revtex4}
\usepackage{latexsym}
\usepackage[dvips]{graphicx}
\usepackage[dvips]{color}
\newcommand{\beq}{\begin{eqnarray}}
\newcommand{\eeq}{\end{eqnarray}}
\newcommand{\eq}{eqnarray}

\newcommand{\al}{{\alpha}}

\newcommand{\ci}{\cite}

\newcommand{\de}{{\delta}}
\newcommand{\De}{\Delta}

\newcommand{\La}{{\Lambda}}

\newcommand{\Om}{{\Omega}}
\newcommand{\pa}{{\partial}}
\newcommand{\no}{{\nonumber}}
\newcommand{\f}{\frac}
\newcommand{\ra}{\rightarrow}

\newcommand{\Ho}{Ho\v{r}ava}

\begin{document}

\preprint{arXiv:2008.06574v6c [hep-th]}
\today\\

\title{Rotating Black Holes in Three-Dimensional {Ho\v{r}ava} Gravity
 {Revisited}
 }

\author{Mu-In Park\footnote{E-mail address: muinpark@gmail.com}}

\affiliation{ Center for Quantum Spacetime, Sogang University,
Seoul, 121-742, Korea }

\begin{abstract}
I 
{revisit} rotating black hole solutions in three-dimensional Ho\v{r}ava gravity with $z=2$
as a simpler set-up of the renormalizable quantum gravity
{\it \`{a} la} 
Lifshitz and DeWitt. The solutions have a curvature singularity at the
origin for a non-vanishing rotation parameter ${\cal J}$, unlike the
black holes in three-dimensional Einstein gravity. For
{\it anti-de Sitter} space, there are black hole event horizons
as usual and the singularity is not naked, in agreement with the cosmic
censorship. {On the other hand, for}
{\it flat} or {\it de Sitter} space, {the earlier solution has also a cosmic-censorship problem because} there are {\it no conventional} black hole horizons as in Einstein gravity,
 other than the usual cosmological horizon for the latter case, so that the
 singularity {\it could} be naked in Ho\v{r}ava gravity. However, {with the help of recent corrections,} I show that the solutions
 have a
 {\it peculiar}
black hole horizon at the origin so that {\it the singularity is not naked
even without the conventional black hole horizons} in
flat or de Sitter case, {due to the Lorentz-violating higher-derivative terms}. On the other hand, I note also that a {\it new ``cosmological" horizon exists even for the flat case}, contrary to the usual wisdom, due to combined effects of the higher derivatives and the angular-momentum barrier. I study {an {\it unified} treatment of} their unusual black hole thermodynamics {for the flat and {de Sitter spaces}, as well as the {anti-de Sitter} space,} {which might be} due to lack of the absolute horizons in the Lorentz-violating gravity.
\end{abstract}



\maketitle

\newpage

\section{Introduction}

{
Several years ago \ci{Park:1207}, I have studied an exact solution for rotating
black holes in three-dimensional Ho\v{r}ava gravity with the dynamical
critical exponent $z=2$ \ci{Soti} as a simpler set-up of the
renormalizable quantum gravity
{\it \`{a} la} 
Lifshitz and DeWitt (HLD) \ci{Lifs,DeWi,Hora}.} The solution has a curvature singularity at the origin $r=0$ for a non-vanishing rotation parameter ${\cal J}$, unlike the absence of the singularity for the black hole solution in three-dimensional Anti-de Sitter (AdS) space \ci{Bana} or the three-dimensional Kerr-de Sitter solution
\ci{Park:9806}, and the ring
singularity for the four-dimensional Kerr black hole in Einstein gravity
\ci{Kerr}. For
AdS space, there are two black hole event horizons $r_{\pm}$ generally as usual where the apparent and Killing horizons coincide and the singularity is not naked, in agreement with the cosmic censorship conjecture \ci{Penr}. On the other hand, for {\it flat} or {\it de Sitter (dS)} space, there are {\it no conventional} black hole horizons as in Einstein gravity, other than the usual cosmological horizon for the latter case, {but those were not considered further because} the curvature singularity {\it could} be naked in three-dimensional Ho\v{r}ava gravity, contrary to the cosmic censorship.

In this paper, I revisit and resolve that issue {of cosmic censorship for flat or dS case,} from the recent corrections in \ci{Park:1207}. The resolution is possible essentially due to the fact that the corrected solution in \ci{Park:1207} shows a {\it peculiar} (non-conventional) black hole event horizon at the origin $r=0$ which coincides with the curvature singularity so that {\it the singularity is not naked even without the conventional black hole horizons} in three-dimensional flat or dS case! I also note that a {\it new ``cosmological" horizon exists even for the flat case}, contrary to the usual wisdom, due to combined effects of the higher derivatives and the angular-momentum barrier. I study their unusual black hole thermodynamics {which might be} due to lack of the absolute horizons in the Lorentz-violating gravity and show that the basic results are unchanged from the earlier work in \ci{Park:1207}.

{The organization of the paper is as follows. In Sec. II, I revisit the previous derivation of rotating black holes in three-dimensional Ho\v{r}ava gravity but with the recent corrections \ci{Park:1207}. In Sec. III, I consider the horizon structure and show that there is a {\it peculiar} black hole event horizon at the origin $r=0$ and the new horizon is crucial to resolve the cosmic censorship problem for flat or dS case in the earlier work \ci{Park:1207}. In Sec. IV, I consider an unified treatment of the black hole thermodynamics for the flat and {dS spaces}, as well as the {AdS} space. In Sec. V, I conclude with some remarks on remaining problems.}

\section{Rotating Black Holes in Three-Dimensional Ho\v{r}ava Gravity  Revisited}

{In this section, for completeness, I revisit the previous derivation of rotating black holes in three-dimensional Ho\v{r}ava gravity but with the recent corrections \ci{Park:1207}, which are crucial to resolve the cosmic censorship issue for the earlier solution of flat or dS case.} In three-dimensional spacetime, gravity becomes enormously simplified. For example, with Einstein-Hilbert action \footnote{However, even in three dimensions, the graviton
modes may exist when higher-derivative terms are present, as in topologically massive gravity
(TMG) \ci{Dese:1982} or new massive gravity (NMG) \ci{Berg:2009}, for example.},
there is no propagating, dynamical degrees of freedom
so that there is no graviton that mediates gravity interactions between massive objects
 \ci{Dese} \footnote{For a comoving frame on  collapsing matters, the (negative) binding energy still exits for a positive Newton's constant, $G_3>0$, as in the conventional four and higher-dimensional black holes \ci{Mann}.}.

However, the system is not too simple to get trivial results only. For example, there exits the black hole solution for three-dimensional  AdS space, known as BTZ solution \ci{Bana}. This suggests that the three-dimensional spacetime could be a good laboratory for studying rotating black holes in Ho\v{r}ava gravity. Actually, it turns out to be the case and one can get the exact solution for rotating black holes in three-dimensional Ho\v{r}ava gravity,
contrary to those
in four dimensions whose exact solutions have not been found yet. In this section, I revisit the previously obtained solution but with the recent corrections in \ci{Park:1207}.

To this ends, I start by considering the ADM decomposition of the metric
\begin{\eq}
ds^2=-N^2 c_\ell^2 dt^2
+g_{ij}\left(dx^i+N^i dt\right)\left(dx^j+N^j dt\right)\,
\end{\eq}
and the three-dimensional renormalizable action with $z=2$,
{\it {\'a} la} Ho\v{r}ara, Lifshitz, and DeWitt \ci{Soti}, which reads
\begin{\eq}
I=\frac{1}{\kappa} \int dt d^2 x \sqrt{g} N
\left( K_{ij}K_{ij}-\lambda K^2 +\xi R +\alpha R^2 -2 \Lambda\right),
\label{action}
\end{\eq}
up to surface terms, where $\kappa \equiv 16 \pi G_3$,
\begin{\eq}
K_{ij}=\frac{1}{2N}\left(\dot{g}_{ij}
-\nabla_i N_j-\nabla_jN_i\right)\
\end{\eq}
is the extrinsic curvature, $R$ is the Ricci scalar of the Euclidean two-geometry, $\lambda,\xi$ are the IR Lorentz-violating parameters, and $\La$ is the cosmological constant. Note that the action (\ref{action}) is general enough since in two-spatial dimensions all curvature invariants can be expressed by the Ricci scalar $R$ due to the identities, $R_{ijkl}=(g_{ik} g_{jl}-g_{il} g_{jk}) R/2,~ R_{ij}= g_{ij} R/2$. However,
{for simplicity}, I do not consider the terms like
$\nabla ^2 R$ \ci{Soti} since the qualitative structure of the
solutions is expected to be similar, as in the four dimensions \ci{Kiri}.
{Here, I do not consider the terms which depend on
$a_i \equiv \pa_i N/N$ and $\nabla_j a_i$ \ci{Blas:2009} either so that Einstein gravity
can be recovered at the low-energy (or  IR) limit: The $a_i$-extended Ho\v{r}ava gravity does
not allow the known solutions in Einstein gravity even in IR because
those extension terms will change the dynamical degrees of freedom and
also the IR as well as UV behaviors a lot from those of the standard
action (\ref{action}) \ci{Donn:2011}, and so the extended case needs a
separate consideration as a different class of gravity theory.
}

Let me now consider an axially symmetric solution with the metric ansatz (I adopt the convention of $c_\ell \equiv 1$, hereafter)
\begin{\eq}
\label{ansatz}
ds^2=-N^2(r) dt^2+\f{1}{f(r)} dr^2
+r^2\left(d \phi+N^\phi(r) dt\right)^2.
\end{\eq}
Substituting the metric ansatz into the action (\ref{action}) gives the
reduced Lagrangian, after angular integration,
\begin{\eq}
\label{eff_action}
{\cal{L}}&=&\frac{2\pi}{\kappa}\frac{N}{\sqrt{f}} \left[\f{f r^3
\left({N^{\phi}}^{'} \right)^2}{2 N^2}-\xi f' +\alpha
\f{f'^2}{r}-2\Lambda r \right]\ ,
\end{\eq}
where the prime $(')$ denotes the derivative with respect to $r$. Note that there is no dependance on $\lambda$ but only on $\xi$ in the Lagrangian, due to the peculiar property $K \equiv g_{ij} K^{ij}=0$ for the ansatz (\ref{ansatz}) \footnote{This gives a peculiar constraint algebra in three dimensions. For the fully {non-linear} constraint analysis, see \ci{Deve:2020}.}.

Varying the metric functions $N$, $N^{\phi}$, and $f$ give the equations of motions as follows,  respectively:
\begin{\eq}
&&-\f{f r^3 ({N^{\phi}}')^2}{2 N^2}-\xi f' +\alpha
\f{f'^2}{r}-2\Lambda r=0,\\
&&\left( \f{\sqrt{f}}{N} r^3 {N^{\phi}}'\right)'=0, \\
&& \left( \f{N}{\sqrt{f}}\right)' \left( 2 \alpha \f{f'}{r}
-\xi\right)+2 \alpha \f{N}{\sqrt{f}} \left(
\f{f''}{r}-\f{f'}{r^2}\right)=0.
\label{eom}
\end{\eq}
{The strategy to solve the coupled non-linear equations is as follows: (i) We first solve (6) for $f$ by expressing $(\sqrt{f}/N)r^3 {N^{\phi}}'={\cal J}$ with
an integration constant ${\cal J}$ which solves (7).
(ii) Then, by plugging the obtained solution for $f$ into (8),
we solve it for $N/\sqrt{f}$.
(iii) Once the solution for
$N/\sqrt{f}$ is obtained, one can solve the previous equation
$(\sqrt{f}/N)r^3 {N^{\phi}}'={\cal J}$ for $N^{\phi}$.
Here, each step depends on the existence of exact solution in the
previous step and it is not a trivial matter to find the
exact solutions for all the steps.
In the earlier published work \ci{Park:1207}, there was an error in identifying the solution for $N$ in the step (ii) and, as the results, $N^{\phi}$ also in step (iii). These corrections do not affect the main conclusion of the work but they are crucial for resolving the naked singularity issue for flat or dS space.}

For arbitrary $\alpha$, $\La$, and $\xi$, the general solution is obtained as \footnote{The solution for $W$ can also be obtained rather easily by observing that (\ref{eom}) can be written as a total derivative form. I thank D. O. Devecioglu for pointing out this. }
\begin{\eq}
f&=&-{\cal M} +\f{b r^2}{2} \left[ 1- \sqrt{a+
\f{c}{r^4}}+\sqrt{\f{c}{r^4}}
ln\left(\sqrt{\f{c}{ar^4}}+\sqrt{1+\f{c}{ar^4}} \right)
\right], \no  \\
\f{N}{\sqrt{f}}&\equiv& W={1}\Big{/}{\sqrt{1+\f{c}{ar^4}}} , \no \\
N^{\phi}&=&-\f{{\cal J}}{2}\sqrt{\f{a}{c}}~ ln
\left[ \sqrt{\f{c}{a r^4}}+\sqrt{1+\f{c}{a r^4}} \right],
\label{f_Horava}
\end{\eq}
where
\begin{\eq}
a \equiv 1+\f{8 \alpha \Lambda}{\xi^2},~b \equiv \f{\xi}{2 \alpha},~ c \equiv \f{2 \alpha
{\cal J}^2}{\xi^2}, \label{abc}
\end{\eq}
and I have set $W(\infty) \equiv 1,~ N^{\phi}(\infty) \equiv 0$ so that the solution approaches to the three-dimensional
{(A)dS
asymptotically}, depending on the sign of $\La$ \ci{Bana,Park:9806,Dese}.
Here, I assume $a>0, c \geq 0$, or equivalently,
$8 \al \La/\xi^2 >-1$ and $\al \geq 0$ so that the metric functions $f, N$,
and $N^{\phi}$ are all real-valued.

For large $r$  and small $\alpha$, one can expand the solution (\ref{f_Horava}) as
\begin{\eq}
f&=&-\f{\Lambda}{\xi} r^2 \left(1-\f{2 \alpha \Lambda}{\xi^2}\right)
-{\cal M} +\f{{\cal J}^2}{4 r^2 \xi} \left(1-\f{4 \alpha
\Lambda}{\xi^2} \right)  -\f{\alpha {\cal J}^4}{24 \xi^3} \f{1}{r^6}
+{\cal O}(\alpha^2, {r^{-10}} ), \no \\
W&=& 1-\f{\alpha {\cal J}^2}{\xi^2}\f{1}{r^4} +{\cal O} ( \alpha^2,{r^{-8}}),
\no \\
N^{\phi}&=&-\f{{\cal J}}{2 r^2} +\f{\alpha {\cal J}^3}{6 \xi^2} \f{1}{r^6}
+{\cal O} (\alpha^2,{r^{-10}} ).
\label{asymp_sol}
\end{\eq}
{Note that the (two) leading terms of $W$ and $N^{\phi}$ are the same as in the earlier work \ci{Park:1207} so that the conserved quantities and their thermodynamics relations in \ci{Park:1207}, which are defined at the asymptotic infinity, are unchanged.}
In the Einstein gravity limit of $\alpha \rightarrow 0$, the solution reduces to
\begin{\eq}
N^2=f=-\f{\Lambda}{\xi} r^2 -{\cal M}+\f{{\cal J}^2}{4
r^2 \xi},~ N^{\phi}=-\f{{\cal J}}{2 r^2},
\label{BTZ}
\end{\eq}
which corresponds (with $\xi=1$) to the BTZ black hole solution for $\La<0, {\cal M}>0$ \ci{Bana}; the three-dimensional Kerr-de Sitter solution ($KdS_3$) for $\La>0, {\cal M}<0$ \ci{Park:9806}; the three-dimensional conical space-time outside of spinning point masses for $\La=0, {\cal M}<0$ \ci{Dese}.
{On the other hand, for the non-rotating limit of ${\cal J} \ra 0$, the solution (\ref{f_Horava}) reduces to the same limit of the GR solution (\ref{BTZ}) but with a modified cosmological constant $\La_{\rm eff}=- \xi b (1-\sqrt{a})/2$.}

The non-vanishing curvature invariants are given by
\begin{\eq}
R&=&-\f{f'}{r}=-b \left(1-\sqrt{a+ \f{c}{r^4}} \right), \no \\
&=&-\f{\xi}{2 \al} \left(1-\sqrt{1+ \f{8 \alpha \Lambda}{\xi^2} } \right)
+\f{{\cal J}^2}{2 \xi \sqrt{1+ \f{8 \alpha \Lambda}{\xi^2} } }\f{1}{r^4} +{\cal O} ({\cal J}^4 r^{-8} ),
\label{R}\\
K^{ij} K_{ij} &=&\f{r^2}{2 W^2} \left( {N^{\phi}}^{'} \right)^2 \no \\
&=&\f{{\cal J}^2}{2 r^4}
\label{KK}
\end{\eq}
and these show the curvature singularities at $r=0$ for ${\cal J} \neq 0 $. This is in contrast to Einstein gravity case $(\al=0, \xi=1)$, where the (covariant) three-curvature scalar $R^{(3)}$ becomes finite, $R^{(3)}=R+K^{ij}
K_{ij}-K^2-f'/r-f''=6 \Lambda$, due to exact concelebration of the unphysical singularities in $R, K^{ij} K_{ij}$, and the remainders.

\section{The Horizon Structure 
and Cosmic Censorship}
{The horizon structure depends on the cosmological constant $\La$. For the AdS case ($\La<0$), I correct the earlier work \ci{Park:1207} based on the corrected solution which is described in Sec. II. For flat and dS case ($\La \geq 0$), which have not be studied before due to the cosmic censorship problems, I show that the corrected solution is crucial to resolve the problem.}

\subsection{The AdS case}

\begin{figure}
\includegraphics[width=8cm,keepaspectratio]{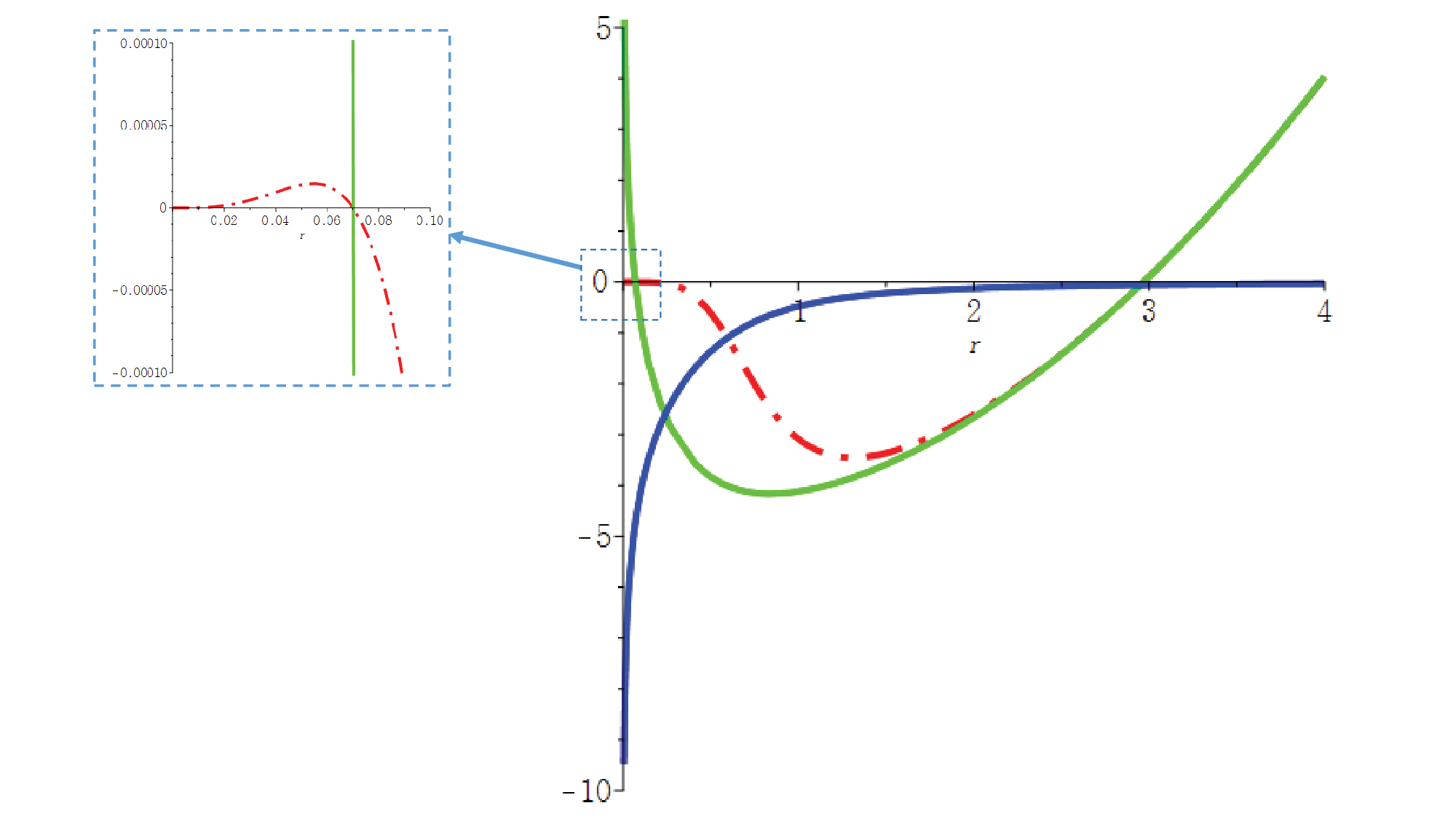}
\includegraphics[width=8cm,keepaspectratio]{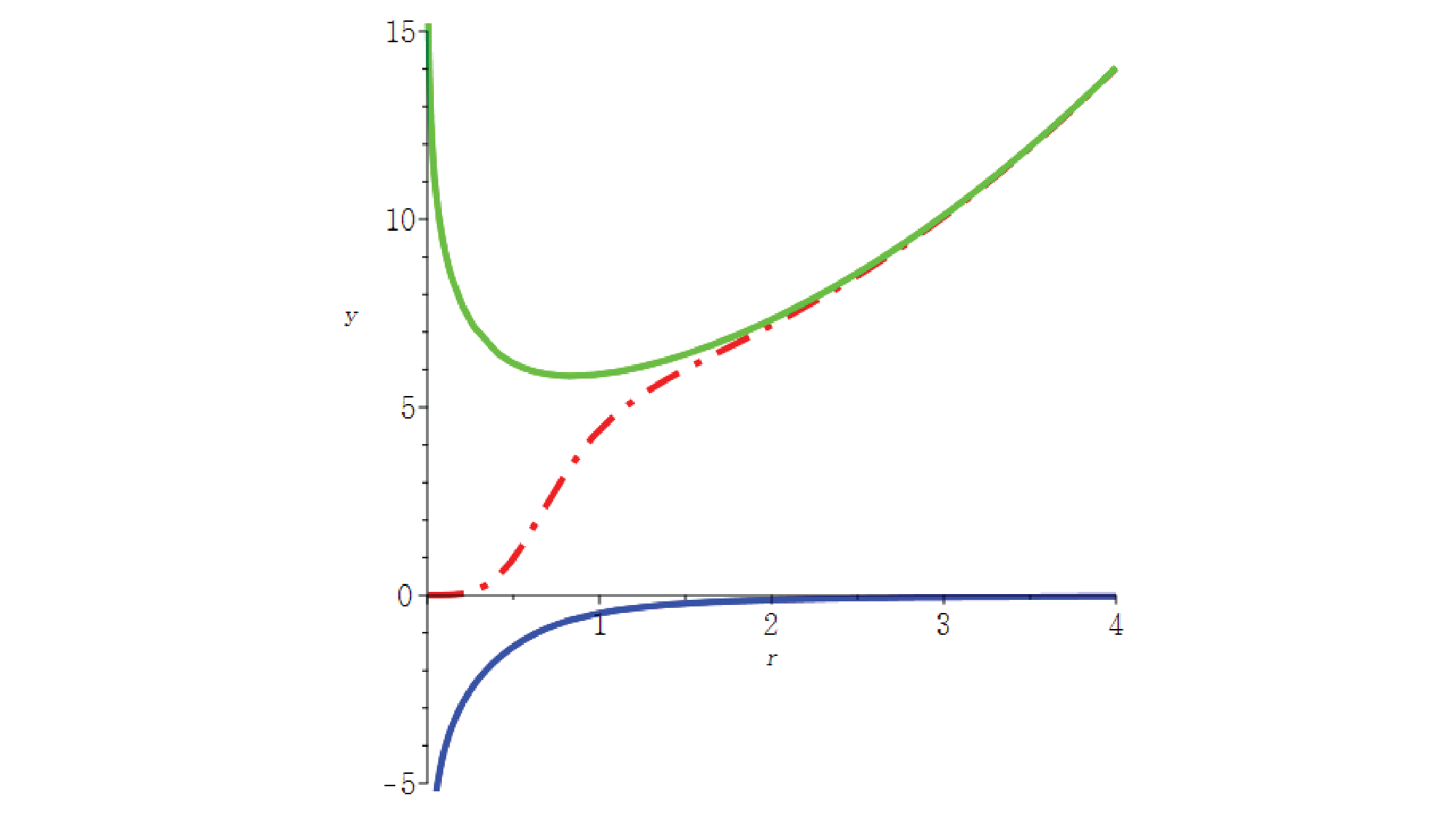}
\caption{Plots of $f(r)$ (green, bright solid), $N^2(r)=W^2 f(r)$ (red, dash-dotted),
$N^{\phi}(r)$ (blue, dark solid) curves for  AdS space with ${\cal M}>0$ (left) and ${\cal M}\leq 0$ (right). Here, I have plotted ${\cal M}=5,-5$ with $\xi=1, \Lambda=-0.5, {\cal J}=1, \alpha=0.1$. For ${\cal M}>0$, there are two horizons, $r_+, r_-$, which are solutions of $f=0,~N^2=0$,
simultaneously, as usual (left). There is another new horizon at the origin $r=0$, where $N^2=0$ but $f \neq 0$. For ${\cal M}\leq 0$, the solution does not have the usual black hole horizons and corresponds to a point particle solution (right).}
\label{fig:N_f_AdS}
\end{figure}

For the AdS space, {\it i.e.}, $\Lambda <0$ and ${\cal M}>0$, the solution (\ref{f_Horava}) has two black hole horizons, $r_+, r_-$, generally where $f$ and $N$ vanish simultaneously so that the apparent and Killing horizons coincide (Fig. \ref{fig:N_f_AdS} (left)) and the curvature singularity at the origin is not naked, in agreement with the cosmic censorship \ci{Penr}.

The Hawking temperature for the black hole horizon, $r_{\!_H}=r_+, r_-$, is given by
\begin{\eq}
T_H&=& {\f{ \hbar \kappa|_{r_{\!_H}}}{2 \pi} }
\no \\
&=&\left.\f{\hbar}{4 \pi} b r_{\!_H} \left(1-\sqrt{a+\f{c}{r_{\!_H}^4}} \right)
\right/\sqrt{1+\f{c}{a r_{\!_H}^4}}
\label{Temp}
\end{\eq}
from { the surface gravity
$\kappa|_{r_{\!_H}}={ (W f')|_{r_{\!_H}}}/{2}$ at the horizon},
as usual. {Here,} the very meaning of the horizon and its Hawking
temperature in Ho\v{r}ava gravity {would not be}
the same as those of Einstein gravity. However, we have used
the same mathematical definition of Killing/apparent horizon and the
Hawking temperature
{as what} the IR, {\it i.e.} lower energy, test particles with the
Lorentz symmetry probe {(for a more detailed discussion, see
Sec. V, Discussion No. 2)}.

\begin{figure}
\includegraphics[width=7cm,keepaspectratio]{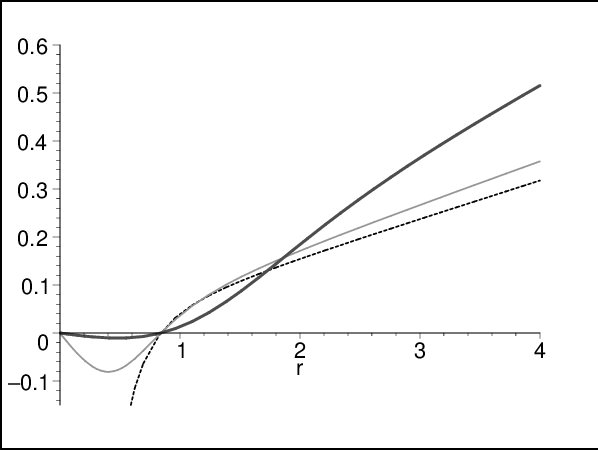}
\qquad
\includegraphics[width=7cm,keepaspectratio]{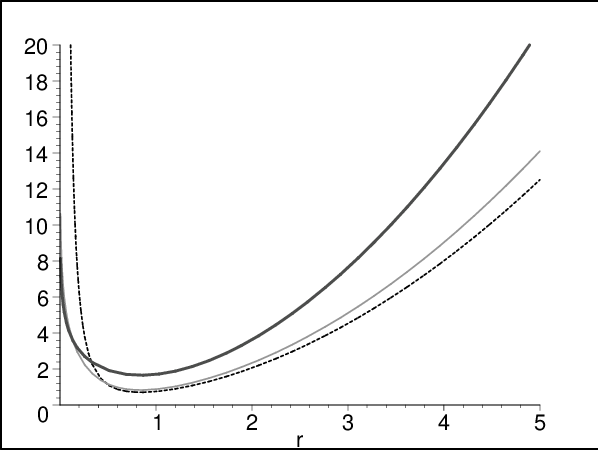}
\caption{Plots of $T_H$ (left) and ${\cal M}$ (right) vs.
$r_{\!_H}$ for AdS space. The two solid curves represent the
three-dimensional rotating Ho\v{r}ava black holes for different
Lorentz-violating higher-derivative coupling $\alpha=0.24,~0.1$ for
the dark and bright curves, respectively, in comparison with the BTZ
case ($\alpha=0$) in the dotted curve. Here, I have considered
$\xi=1,\Lambda=-0.5,{\cal J}=1$, and $\hbar \equiv 1$.
}\label{fig:Temp_AdS}
\end{figure}

For a non-vanishing $c$ or ${\cal J}$, the temperature vanishes in the usual extremal black hole limit, where the inner horizon $r_-$ meets the outer horizon $r_+$, {\it i.e.}, a degenerate horizon exists at
\begin{\eq}
r_{H}^{*}=\left( \f{c}{1-a}\right)^{1/4}
=\left( \f{{\cal J}^2}{-4 \La}\right)^{1/4}
\end{\eq}
(Fig. \ref{fig:Temp_AdS} (left)) and the integration constant,
\begin{\eq}
\label{bare_mass}
{\cal M} =\f{b r_{\!_H}^2}{2} \left[ 1- \sqrt{a+
\f{c}{r_{\!_H}^4}}+\sqrt{\f{c}{r_{\!_H}^4}}~
{ln}\left(\sqrt{\f{c}{a r_{\!_H}^4}}+\sqrt{1+\f{c}{a r_{\!_H}^4}} \right)
 \right]
\end{\eq}
gets the minimum ${\cal M}^*$ (Fig. \ref{fig:Temp_AdS} (right)),
\begin{\eq}
{\cal M}^*&=&\f{ b \sqrt{c}}{2} ~ln \left(\f{\sqrt{1-a}+1}{\sqrt{a}} \right) \no \\
&=&\f{ \xi }{4 \al} \sqrt{\f{2 \al {\cal J}^2 }{\xi^2}} ~ln \left(\f{\sqrt{\f{-8 \al \La}{\xi^2}}+1}{\sqrt{\f{1+8 \al \La}{\xi^2}}} \right).
\end{\eq}
{Here,} it is interesting to note hat $r^*_H$ has no
{\it explicit} $\alpha$-dependence and has the same expression as in Einstein gravity,
in contrast to the explicit $\alpha$-dependence in ${\cal M}$ and its minimum value ${\cal M}^*$,
as can be seen also in Fig. 2.

In addition to these conventional black hole horizons, $r_+, r_-$, remarkably, one can also find that there is another peculiar horizon at the origin $r_{--}=0$, if $c \neq 0, {\it i.e.}$, $\al \neq 0$ and ${\cal J} \neq 0$, where $N^2=W^2 f=0$ but $f \neq 0$, which corresponds to an {\it apparent} horizon but not a {\it Killing} horizon, from the behaviors of the metric near the origin \footnote{Note that the limiting behavior of  $x^n~ ln x \ra 0^{-}$ as $x \ra 0^+$, or equivalently $ln x /x^n \ra 0^{+}$ as $x \ra \infty$, for $n>0$. The latter limit is usually summarized as ``{logarithms grow more slowly than any power or root of $x$}". } (Fig. \ref{fig:N_f_AdS} (left)),
\begin{\eq}
f&=&-b \sqrt{c}~ ln r-{\cal M} +\f{b \sqrt{c}}{2}
\left( ln \left(2 \sqrt{ \f{c}{a}} \right)-1\right)+\f{b }{2}r^2+{\cal O}(r^4), \no  \\
N^2&=&\left( \f{a}{ c} \right) \left[- {\cal M} + \f{b \sqrt{c}}{2} \left(  ln \left(2 \sqrt{ \f{c}{a}}\right)-1\right) \right] r^4-\f{ab }{\sqrt{c}}~r^4~ ln r +\f{ab}{2 c}~ r^6 +{\cal O}(r^8~ ln r, r^8).
\label{f_near_origin}
\end{\eq}
{Here,} the apparent
horizon is defined by the null hypersurface $g^{\mu \nu}(\pa_\mu r)(\pa_\nu r) = N^2 = 0$, whereas the Killing horizon by the surface where the norm of the Killing vector $\chi = \pa_t + \Om_H \pa_\phi$ vanishes, {\it i.e.}, $\chi^2= g_{tt}-(g_{t\phi})^2/g_{\phi \phi} = -f^2= 0$ with the angular velocity of the horizon $\Om_H = -(g_{t\phi}/g_{\phi \phi})|_{H}$. In the conventional stationary black holes, these two horizons coincide; for the black holes in (three-dimensional) non-commutative space, there is a splitting of the two horizons due to the non-commutativity \ci{Kim:2007}. But for the new horizon in our three-dimensional Ho\^{r}ava gravity, there exists only the apparent horizon
{\it without} its paired Killing horizon.

The existence of a new horizon at the origin can also be checked by the vanishing of temperature $T_H$ (Fig. \ref{fig:Temp_AdS} (left)) as
\begin{\eq}
T_H =-b \sqrt{a}~ r_{\!_H} + b \sqrt{\f{a}{c}} ~r_{\!_H}^3 + {\cal O} (r_{\!_H}^5)
\end{\eq}
near $r_{\!_H}=0$, which implying another ``degenerate" horizon at the origin, though the ``negative" Hawking temperature for the inner horizon $r_-$ would not make sense along the way to the origin $r_-=0$, due to the lower bound of the mass spectrum \ci{Park:0602}. Actually, as ${\cal M}$ increases, the inner horizon $r_-$ approaches to the origin where the peculiar horizon $r_{--}=0$ is located and finally $r_-$ coincides with $r_{--}=0$ at ${\cal M}\ra \infty$ limit (Fig. 2 (right)), which can be seen analytically as
\begin{\eq}
{\cal M}=\f{b \sqrt{c}}{2}
\left( ln \left(2 \sqrt{ \f{c}{a}} \right)-1\right)+\f{b }{2}r_{\!_H}^2-b \sqrt{c}~ ln r_{\!_H} +{\cal O}(r_{\!_H}^4).
\end{\eq}

However, the new horizon at the origin would {\it not} have any important effect for the outsider
observer since the usual black hole horizons are already formed and hides the new horizon at
the origin. It could have an important effect
when there is no conventional black hole horizons at finite radius, like  the case of ${\cal M}\leq0$ (Fig, 1 (right)), which corresponds to a point particle
solution \ci{Park:0602}.
However, even in this case, even though there a curvature singularity at the origin, it is not naked due to the new {\it point-like} (apparent) horizon which coincides with the location of the singularity!

\subsection{The dS case}

For the dS space, {\it i.e.}, $\La>0$, the solution (\ref{f_Horava}) has no conventional black hole horizon but the usual cosmological horizon, $r_{++}$, where $N^2$ and $f$ vanish simultaneously, regardless of the sign of ${\cal M}$ (Fig. 3): In the case of ${\cal J} \neq 0$, there is no constraint on ${\cal M}$ for the existence of the cosmological horizon, ranging from $-\infty$ to $+\infty$, whereas in the case of ${\cal J}=0$, there is no conventional cosmological horizon for ${\cal M}>0$. However, as in the AdS case, a peculiar (apparent) horizon $r_{--}=0$, where
$N^2=0$ but $f \neq 0$ generally, exists but now it becomes the unique
``black hole" horizon which
hides the singularity at $r=0$:
{Near} the horizon $r_{--}=0$, the escape time for the radial null signals from the horizon is $t=\int^r_0 dr /\sqrt{N^2 f}\sim \int^r_0 dr /{r^2 |ln r|}=Ei(lnr)$ which becomes {\it infinite} as $r \ra 0$ so that any signal from the singularity can not be observed \footnote{Here, the {\it exponential integral} $Ei(x)$ is defined as $Ei(x)=-\int^{\infty}_{-x}dt~ e^{-t} t^{-1} $, for real non-zero values of $x$.}; {this property supports our calling $r_{--}=0$ as a horizon for an outside observer, though it cannot be traversed by
any causal curve.}

The curvature singularity at the origin $r=0$, for the non-vanishing $\al$ and the rotation parameter ${\cal J}$, is not naked by the point-like horizon which is degenerated at the same location so that the cosmic censorship is satisfied in an interesting way. This is in contrast to the non-rotating case in the \Ho~gravity solution (\ref{f_Horava}) or $KdS_3$ in Einstein gravity \ci{Park:9806}, where there is neither $r=0$ singularity in (\ref{R}), (\ref{KK}) nor black hole horizons so that the cosmic censorship conjecture is trivially satisfied.

On the other hand, the Hawking temperature for the cosmological horizon, $r_{\!_C}=r_{++}$, in dS space is given by
\begin{figure}
\includegraphics[width=8cm,keepaspectratio]{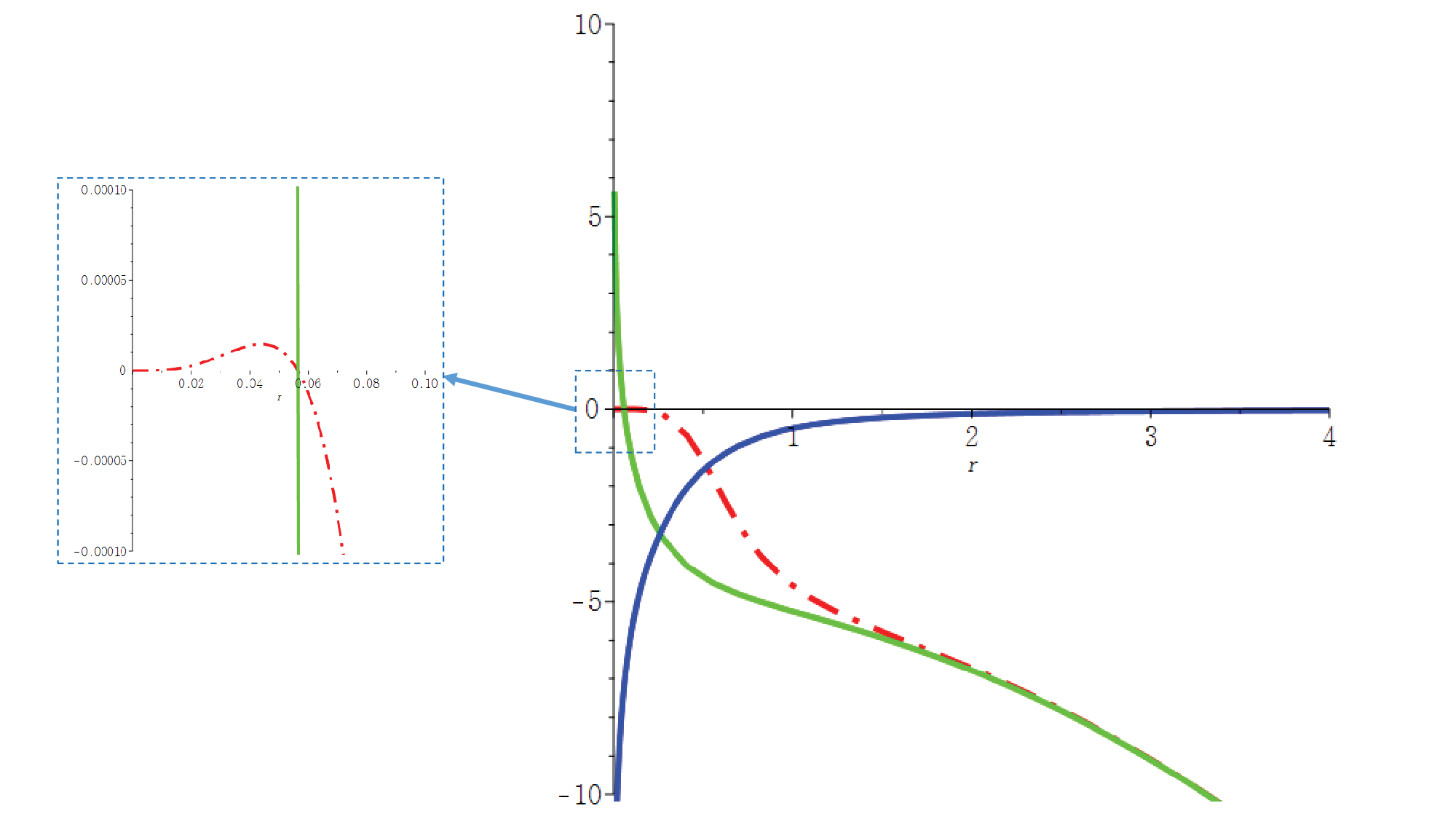}
\includegraphics[width=8cm,keepaspectratio]{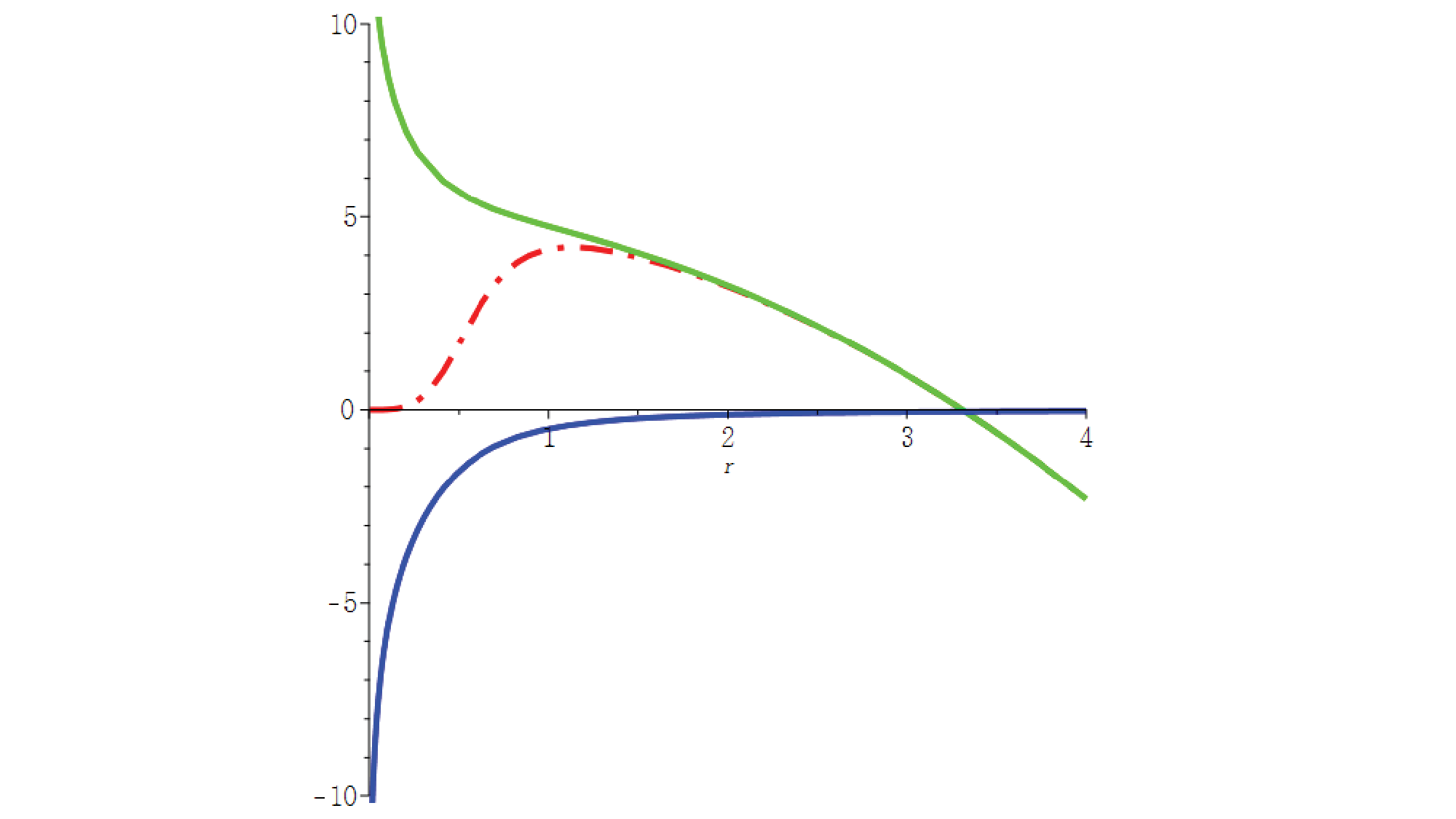}
\caption{Plots of $f(r)$ (green, bright solid), $N^2(r)=W^2 f(r)$ (red, dash-dotted),
$N^{\phi}(r)$ (blue, dark solid) curves for dS space with ${\cal M}>0$ (left)
and ${\cal M}\leq 0$ (right). Here, I have plotted ${\cal M}=5,
-5$, respectively, with $\xi=1, \Lambda=0.5, {\cal J}=1, \alpha=0.1$. There is no conventional black hole horizons but just the usual {\it cosmological} horizon $r_{\!_C}=r_{++}$, which is the solution of $f=0,~N^2=0$, simultaneously. However, there is a peculiar (apparent) {\it black hole} horizon at the origin $r_{--}=0$, where $N^2=0$ but $f \neq 0$. }
\label{fig:N_f_dS}
\end{figure}
\begin{\eq}
T_C&=&\f{ \hbar | W f'|_{r_{\!_C}}}{4 \pi} \no \\
&=&\f{\hbar}{4 \pi} \left|  r_{\!_C} \left(1-\sqrt{a+\f{c}{r_{\!_C}^4}} \right) \right| \left/\sqrt{1+\f{c}{a r_{\!_C}^4}} \right.~~,
\label{Temp_dS}
\end{\eq}
following the usual convention of the {\it positive} Hawking temperature \ci{Gibb,Park:9806,Park:0705} (Fig. 4 (left)) and the integration constant ${\cal M}$ is given by the same formula as in (\ref{bare_mass}) by replacing $r_{\!_H}$ with $r_{\!_C}$ (Fig. 4 (right)). The existence of the two extremal points in the Hawking temperature implies two phase transitions between the stable horizons and the evaporating horizons at
\begin{\eq}
r_{\!_{C1}}&=&\left\{ \f{c}{a (a-1)} \left[ 2-a+ \De +\De^{-1} \right]\right\}, \\
r_{\!_{C2}}&=&\left\{ \f{c}{a (a-1)}
\left[ 2-a-\f{1}{2}(\De +\De^{-1})-\f{\sqrt{3} i}{2} (\De-\De^{-1}) \right]
\right\}^{1/4},
\end{\eq}
where
\begin{\eq}
\De \equiv\left[1-2(a-1)^2 + a (a-1) \sqrt{2(a-2)} \right].
\end{\eq}
\begin{figure}
\includegraphics[width=7cm,keepaspectratio]{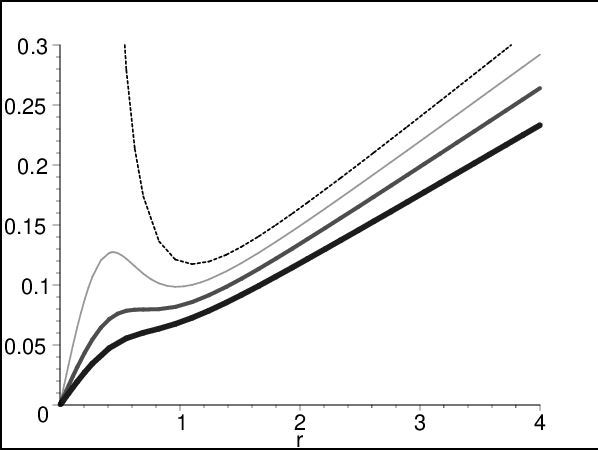}
\qquad
\includegraphics[width=7cm,keepaspectratio]{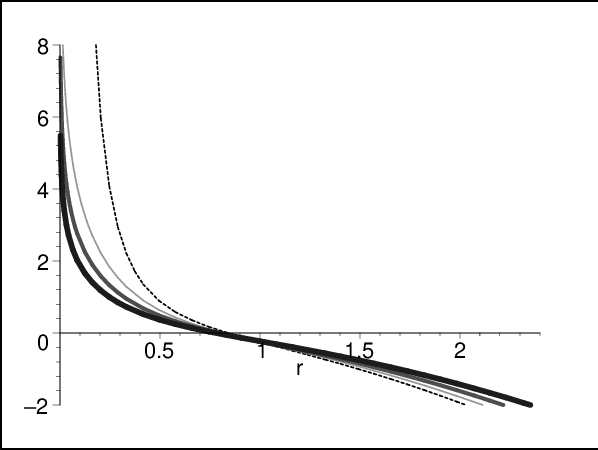}
\caption{Plots of $T_C$ (left) and ${\cal M}$ (right) vs.
$r_{\!_C}$ for dS space. The three solid curves represent the
rotating Ho\v{r}ava-de-Sitter solutions for different
Lorentz-violating higher-derivative coupling $\alpha=0.5,0.25,0.1$
(bottom to top (the left part for ${\cal M}$))
in comparison with the $KdS_3$
case ($\alpha=0$) in the dotted curve. Here, I have considered
$\xi=1,\Lambda=0.5,{\cal J}=1$, and $\hbar \equiv 1$. The two extremal points at $r_{\!_{C1}}$ and $r_{\!_{C2}}$ $(r_{\!_{C1}}>r_{\!_{C2}})$ in the temperature imply two phase transitions between the stable horizons and the evaporating (unstable) horizons.}\label{fig:Temp_dS}
\end{figure}
As the observers in the interior region of the rotating dS space, $r \leq r_{\!_C}$,
detect an isotropic background of thermal radiations with the Gibbons-Hawking temperature
$T_C$ (\ref{Temp_dS}) \ci{Gibb} at the expense of the horizon area,
$T_C$ goes down for the case of $r_{\!_C} \geq r_{\!_{C1}}$, but $T_C$ goes up for the case of $r_{\!_C} < r_{\!_{C1}}$, up to the radius $r_{\!_{C2}}$ of the local maximum in $T_C$; there is a phase transition between the large stable horizons and the
small evaporating (unstable) event horizons at the critical point $r_{\!_C}=r_{\!_{C1}}$. This is
 similar to the case of $KdS_3$ solution in Einstein gravity but
different from the previously studied AdS space black holes in \Ho~gravity \ci{Park:1207} as well as the BTZ black hole in Einstein gravity \ci{Bana}, which are always stable. As the cosmological horizon $r_{\!_C}$ shrinks smaller than $r_{\!_{C2}}$ by the further thermal radiation, $T_C$ goes down again up to the point of vanishing cosmological horizon $r_{\!_C}=0$ so that there is another stable phase for $0 \leq r_{\!_C} < r_{\!_{C2}}$. The existence of a new stable phase is the genuine effect of the higher-derivative terms in \Ho~gravity which persist until a critical coupling $\al_c=\xi^2/8 \La$, {\it i.e.,} $a=2$ so that $\De=-1$ ($\al=0.25$ in Fig. 4 (left)) is being reached.

Moreover, the vanishing Hawking temperature $T_C$ in Fig. 4 (left) as
\begin{\eq}
T_C =\left| -b \sqrt{a}~ r_{\!_C} + b \sqrt{\f{a}{c}} ~r_{\!_C}^3 + {\cal O} (r_{\!_C}^5) \right|
\label{T_C_zero}
\end{\eq}
when the cosmological horizon $r_{\!_C}=r_{++}$ approaches the origin $r_{\!_C}=0$
implies
 the merging of the two horizons $r_{++}$ and $r_{--}=0$, similarly to  AdS case: {For an observer} in AdS space sitting
 at the interior region of $0<r<r_-$, the inner horizon $r_-$ would be considered as a
 cosmological horizon so that the {\it negative} Hawking temperature $T_-$ for the inner
 horizon $r_-$ may be interpreted as the {\it positive} Hawking $T_C \equiv |T_-|$ for the ``inner cosmological horizon" $r_{\!_C} \equiv r_-$
 (cf. \ci{Cvet:2018}), for the consistency with the convention of the {\it outer} cosmological
 horizon $r_{++}$; then, the merging process of the inner cosmological horizon
 $r_-$ and the point-like horizon $r_{--}=0$ is exactly corresponding to
 that of the outer cosmological horizon $r_{++}$.

\subsection{The flat case}

For the flat space, {\it i.e.}, $\La=0$, the solution (\ref{f_Horava}) shows the most dramatic case. The peculiar property of this solution is that, {\it regardless of the absence of the cosmological constant $\La$, there is also another ``cosmological" horizon for ${\cal M}>0$},
contrary to the usual wisdom \footnote{It seems that the existence of a cosmological horizon
even for the three-dimensional flat Einstein gravity $(\al=0)$ has not been well recognized until
recently \ci{Corn}; actually, it could have been recognized earlier since the solution is just at
the border of the BTZ solution for $\La <0, {\cal M}>0$ \ci{Bana} and $KdS_3$ solution for $\La>0, {\cal M}>0$
\ci{Park:9806}. But in this case, there is neither the point horizon nor the curvature
singularity at $r=0$. More recently, similar atypical asymptotics have been
discovered also in the context of {\it Einstein-aether} theory, {or a low-energy
limit of Ho\v{r}ava gravity}
\cite{Soti:2014}. },
as well as the point-like horizon at $r=0$, if $c$ does not vanish, {\it i.e.}, $\al \neq 0$ and ${\cal J}\neq 0$ (Fig. 5). This would be due to the combined effects of the (repulsive) nature of the higher derivatives and the angular-momentum barrier, as can be checked analytically in (\ref{f_near_origin}) also. Due to the point-like horizon at $r=0$, the curvature singularity at the origin is not naked as in the dS case. But, the difference from the dS case is that the new cosmological horizon moves to the spatial infinity as ${\cal M}$ approaches to zero and disappears as ${\cal M}$ becomes negative in the flat case. This is a generic behavior of three-dimensional gravity, without much dependence on the higher curvature terms in \Ho~gravity (Fig. 6). On the other hand, the case of ${\cal M}<0$ corresponds to the point particle solution in \Ho~gravity \footnote{In the Einstein gravity limit $\al \ra 0$ \ci{Dese,Carl:1998},
the mass and angular momentum are given by
$m = \f{1-\sqrt{-{\cal M}}}{4G},
j=\f{{\cal J}}{8G \sqrt{-{\cal M}}},
$
respectively. It would be an interesting problem to identify those for $\al \neq 0$ also
(cf \ci{Bell:2019}).
}.


The integration parameter ${\cal M}$ and the Hawking temperature $T_C$ can be obtained by setting $a=1$, {\it i.e.,} $\La=0$ in (\ref{bare_mass}) and (\ref{Temp_dS}), respectively (Fig. 7). In contrast to the dS case, there is only one extremal point of the temperature with the maximum $T_{max}=\sqrt{2}(2 \sqrt{3}-3)^{3/4} b c^{1/4}/8 \pi$ and this persists for any coupling
as far as the cosmological horizon exists, {\it i.e.,} ${\cal M}>0$.
So, there is only one phase transition between the large evaporating and the small stable (cosmological) horizons at $r=r_{\!_{C2}}=[(2/\sqrt{3}-1)c ]^{1/4}/2$ with $\De=1$.

\begin{figure}
\includegraphics[width=8cm,keepaspectratio]{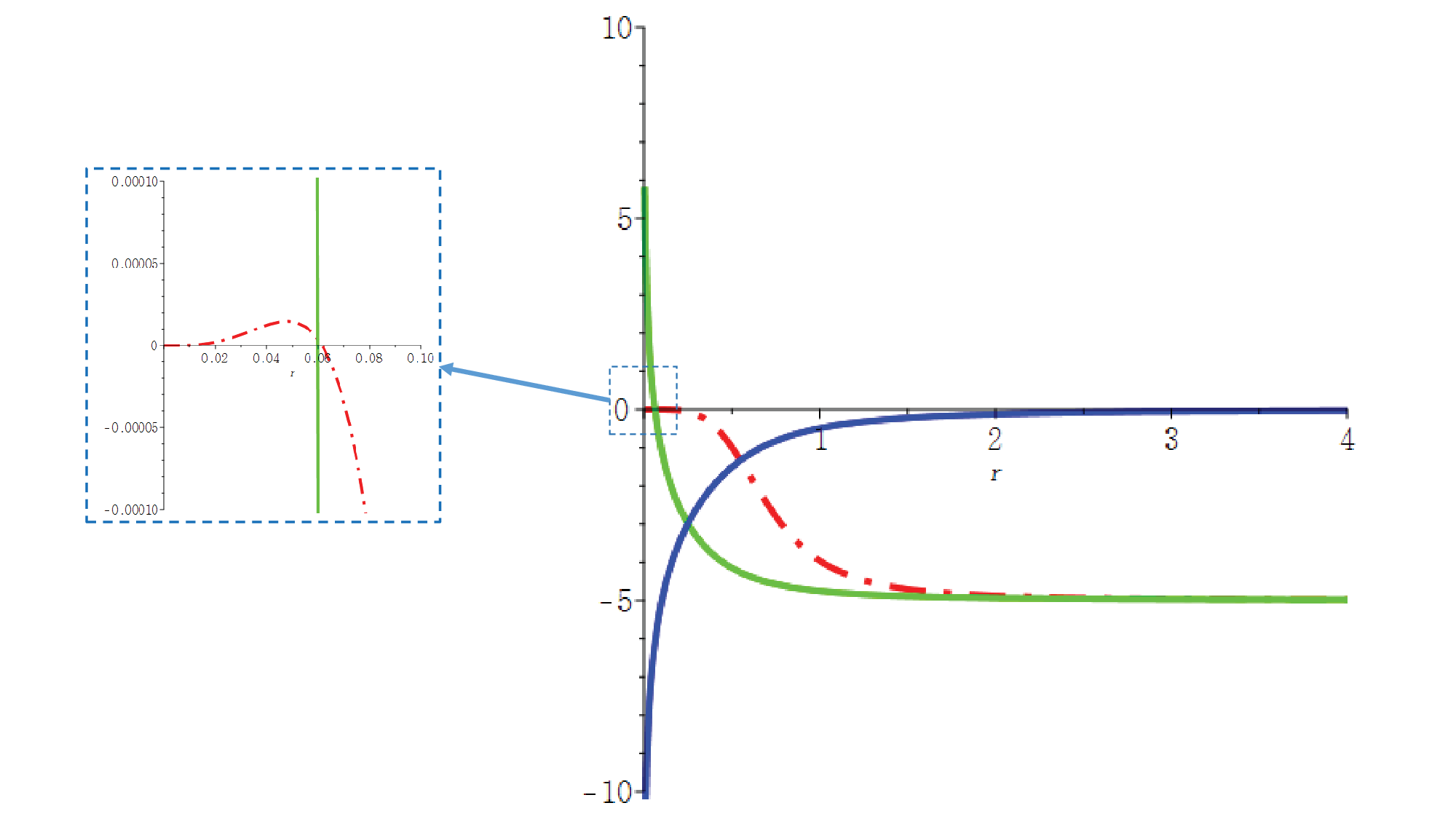}
\includegraphics[width=8cm,keepaspectratio]{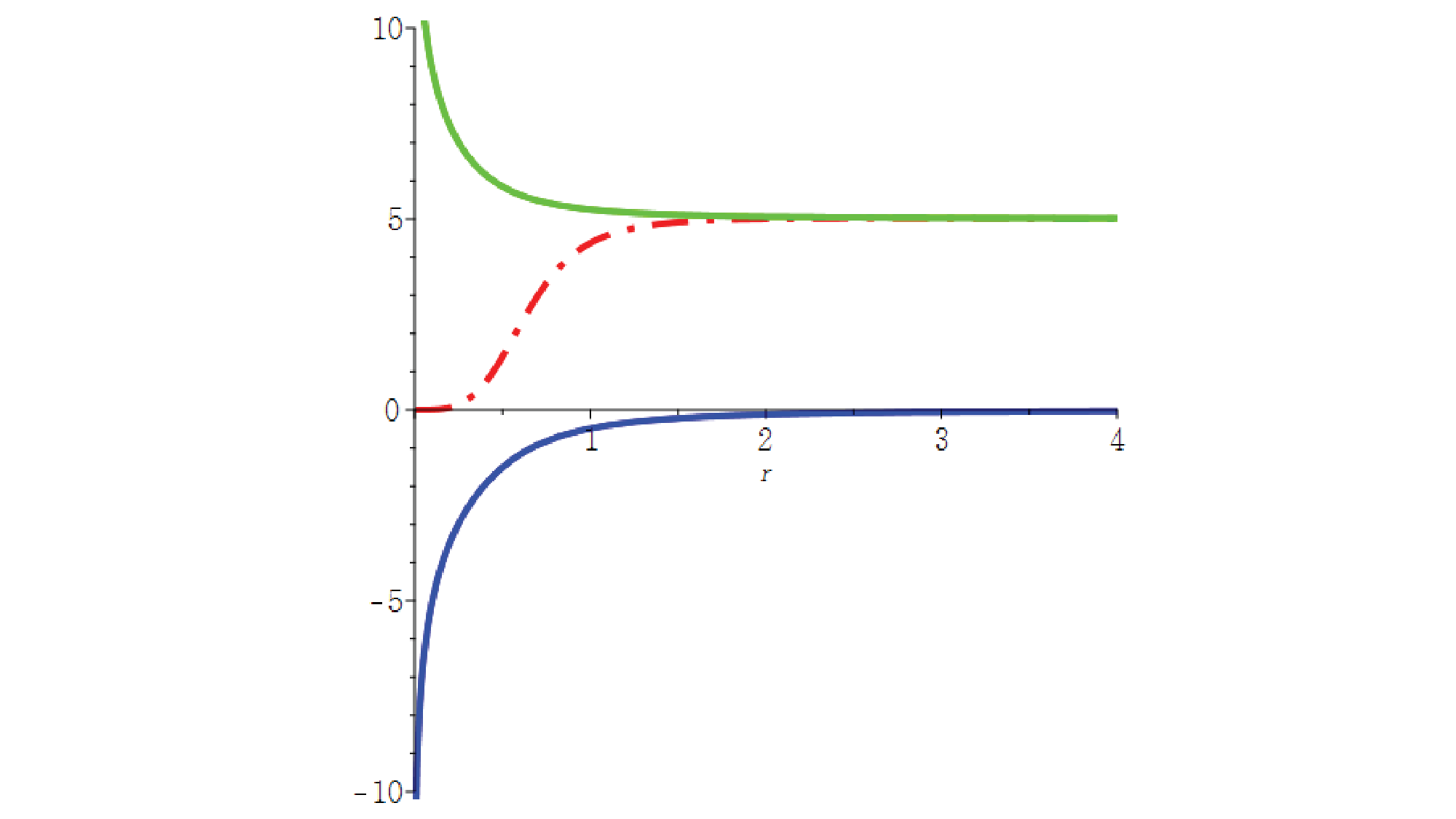}
\caption{Plots of $f(r)$ (green, bright solid), $N^2(r)=W^2 f$ (red, dash-dotted),
$N^{\phi}(r)$ (blue, dark solid) curves for flat space with ${\cal M}>0$ (left, center),
${\cal M}<0$ (right). (The plotted parameters, except $\La=0$, are the same as in Fig. 3)
For ${\cal M}>0$, there is one {\it cosmological} horizon  as well as a peculiar {\it black hole}
horizon at the origin, as in the dS case. For ${\cal M}<0$, there is no {\it conventional}
cosmological/black hole horizon but just the point particle solution remains. }
\label{fig:N_f_flat}
\end{figure}

\begin{figure}
\includegraphics[width=6cm,keepaspectratio]{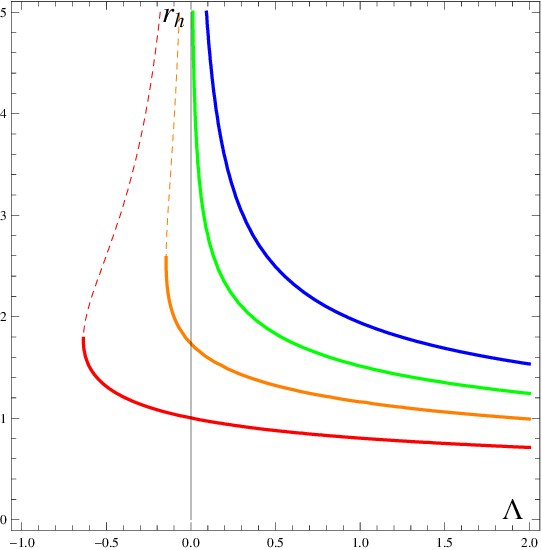}
\caption{Numerical plots of $r_h \equiv (r_{\!_H}, r_{\!_C})$ vs. $\La$ for different masses, ${\cal M}=5,2,0,-2$ (left to right). Depending on the values of ${\cal M}$ and $\La$, there are two branches of solutions. For the first branch (left two curves), the solutions have two (black hole) horizons (dotted lines for outer horizons and solid lines for inner horizon) for AdS ($\La<0$) and their inner horizons (only) remain for flat or dS ($\La \geq 0$). In the latter case, the outer horizons in AdS case moves to infinity and the inner horizons become the cosmological horizons in flat or dS case. For the second branch (right two curves), there is no corresponding solution in AdS or flat but exists only in dS case. Here, I have considered $\xi=1, {\cal J}=5, \alpha=0.1$. }\label{fig:Temp_flat}
\end{figure}

\begin{figure}
\includegraphics[width=7cm,keepaspectratio]{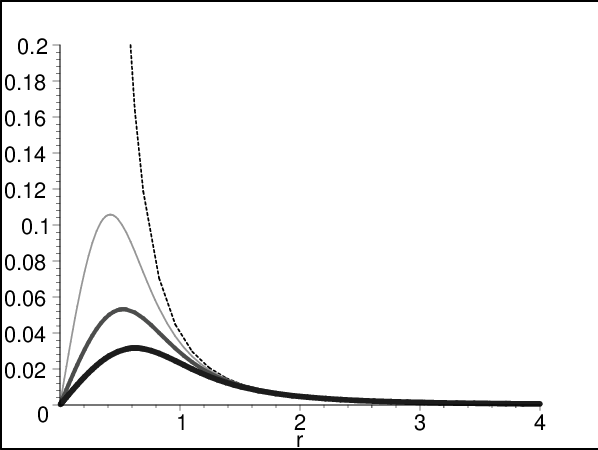}
\qquad
\includegraphics[width=7cm,keepaspectratio]{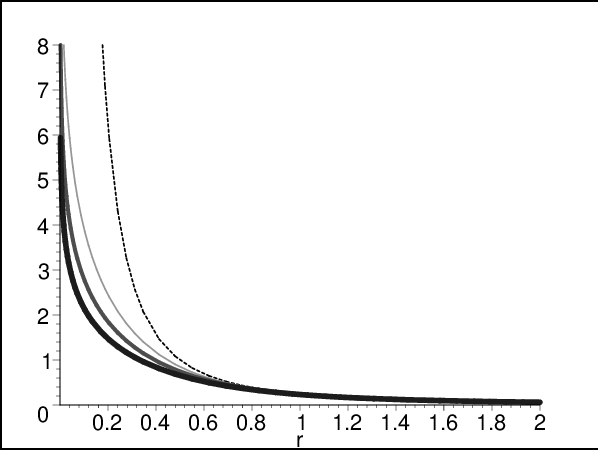}
\caption{Plots of $T_C$ (left) and ${\cal M}$ (right) vs.
$r_{\!_C}$ for flat space. (The plotted parameters are the same as in Fig. 4). There is only one extremal point in the temperature which implies one phase transition between the  (large) unstable cosmological horizon and the (small) stable  horizon.}\label{fig:Temp_flat}
\end{figure}

\section{The Mass, Angular Momentum, and Their Unusual Thermodynamical Law:
An Unified Treatment
}
{The leading terms of the corrected solution (\ref{asymp_sol}) are the same as in the earlier work \ci{Park:1207} and so the conserved quantities and their thermodynamics relations, which are defined at the asymptotic infinity, are unchanged. In this section, I present an {\it unified} treatment of the horizon thermodynamics for the flat and dS cases, as well as the AdS case in previous work \ci{Park:1207}.}

There are some subtleties in defining the conserved quantities when there is the cosmological horizon, partially due to the absence of an asymptotic observer. However, due to some recent developments on the entanglements \ci{Isra}, the extension of the usual methodology to the hypothetical observer in the causally-disconnected regions could be more than just a mathematical apparatus \ci{Park:1207,Park:0705,Stro}. Here, I consider and extend the canonical methodology for computing the conserved mass and angular momentum of the rotating solution (\ref{f_Horava}) for the black hole as well as the cosmology solution, based on the boundary variation of the action. To this ends, I start by considering the variation of the total action $I_{total}=I+B$ with boundary terms $B$ at the infinity $r=\infty$, which may be space-like or time-like, such that the boundary variation $(\de I)(\infty)$ is canceled by $\de B$
and there remain only the bulk terms in $\de I_{total}$, which vanish
when the equations of motions hold. Then, for the class of fields that
approach our solution (\ref{f_Horava}), or more generally the asymptotic form (\ref{asymp_sol}) al least, at $r=\infty$, one finds
\begin{\eq}
B=\pm(t_2-t_1) \left[-W(\infty) M+N^{\phi} (\infty) J \right],
\end{\eq}
which defines the canonical mass and angular momentum
\begin{\eq}
M=\pm\f{2 \pi \xi \sqrt{a}}{\kappa} {\cal M}, ~J=\pm\f{2 \pi \xi }{\kappa}
{\cal J},
\end{\eq}
as the conjugates to the asymptotic displacements $N(\infty)$ and
$N^{\phi}(\infty)$, respectively, when kept as independent
parameters. Here, the upper $(+)$ and lower $(-)$ signs correspond to the AdS and dS/flat cases, respectively, which may be traced back their origins to the opposite role of space and time coordinates on the space-like (AdS case) and the time-like (dS/flat case) boundaries \ci{Brow}. With the above choice of signs, one can now get the same mass and angular momentum as in the literatures for the dS and flat cases \ci{Park:0705,Stro}, as well as the AdS case \ci{Park:1207}.

Even though the conserved mass and angular momentum can be obtained, their thermodynamical law for Lorentz-violating black holes has not been well established yet. Actually, previously for the AdS case, it has been shown that the conventional first law of black hole thermodynamics with the usual Hawking temperature and chemical potential does not work \ci{Park:1207}, and this is fundamentally unchanged even for the dS and flat cases also. To see this, let me consider the variation of the mass $M$ as a function of $J$
and $r_{\!_h}\equiv (r_{\!_H}, r_{\!_C})$, which can be written as
\begin{\eq}
dM =A d r_{\!_h}+\Omega_h dJ  \label{first:0}
\end{\eq}
with the chemical potential $\Omega_h \equiv -N^{\phi}|_h$ and
\begin{\eq}
A \equiv \pm \f{\pi \xi^2}{\kappa \alpha} r_{\!_h}  \sqrt{a} \left(
1-\sqrt{a+\f{c}{r_{\!_h}^4}} \right) \label{A,B}
\end{\eq}
(the upper and lower signs for AdS and dS/flat cases, respectively).

Then, in order to see whether the black hole entropy can be defined through the first law of thermodynamics in the
conventional form
\begin{\eq}
d M=T_{h} d S+\Omega_h dJ
\label{first:1}
\end{\eq}
with the usual Hawking temperature $T_H$ of (\ref{Temp}) for the AdS case,
or $T_C$ of (\ref{Temp_dS}) for the dS/flat case,
let me define the black hole entropy function $S\equiv S(r_{\!_h},J)$, as a function of $r_{\!_h}$ and $J$, with
\begin{\eq}
dS \equiv \partial_{r_{\!_h}} S~ d r_{\!_h} +\partial_{J} S ~dJ.
\end{\eq}
Then, from (\ref{first:0}),
(\ref{A,B}), and (\ref{first:1}), one can find
\begin{\eq}
\partial_{r_{\!_h}} S=\f{A}{T_h},~
\partial_{J} S=0
\end{\eq}
but
\begin{\eq}
\partial_{J}
\partial_{r_{\!_h}} S- \partial_{r_{\!_h}} \partial_{J} S
=\f{\pm 8 \pi \alpha {\cal J}}{\sqrt{a} r_{\!_h}^4 \sqrt{1+c/a r_{\!_h}^4}},
\end{\eq}
which shows that {\it the (horizon) entropy is not integrable} with the
non-relativistic higher-curvature corrections ($\alpha \neq 0$) for
rotating black holes. This proves that the
entropy can {\it not} be defined in the conventional form of the
first law of thermodynamics with the usual Hawking temperature and
chemical potential, for the generic black holes or cosmology solutions with a rotation.

\section{Concluding Remarks }

In conclusion, I have {revisited}
rotating black hole solutions in
the three-dimensional Ho\v{r}ava gravity
for flat or (A)dS case {with
the recent corrections \ci{Park:1207}, and found
a peculiar point-like horizon at the origin, even without the
conventional black hole horizons. The corrections are crucial to
resolve the naked singularity problems in the earlier work since
the curvature singularity at the origin is not naked even for the flat
and dS cases. So I have found
a} \Ho~gravity generalization of the known solutions for the $KdS_3$ and flat solutions{, as well as $BTZ$ solution, in three-dimensional} Einstein gravity. I have also found that a new ``cosmological" horizon exists even for the flat case, contrary to the usual wisdom, due to combined effect of the higher derivatives and the angular-momentum barrier. I have studied {an {\it unified} treatment of} their unusual black hole thermodynamics {for the flat and {dS} spaces, as well as the {AdS} space in previous work \ci{Park:1207},}
{which might be} due to lack of the absolute horizons in the Lorentz-violating gravity and shown that the basic results are unchanged from the earlier work. {Several further remarks about remaining problems are in order.}\\

1. The existence of the point-like horizon at the origin resembles the case of black ``plane" solutions in four-dimensional \Ho~gravity, though the details of horizon structures are different \ci{Argu}. The point-like horizon in three-dimensional \Ho~gravity might be due the lower-derivative equations of motion for $N$ in (\ref{eom}). It would be interesting to see whether the introduction of higher-derivative terms, like the $\nabla^2 R$ term, which has not been considered in this paper, could make the point horizon at $r=0$ to be {\it expanded} so that the more sizable black hole may be obtained \ci{deBu,Park:0801}.\\


{2.} {Regarding the Hawking radiations in Lorentz-violating gravities, there
have been some controversial results. For example, for $(3+1)$-dimensional static
black holes in the Einstein-aether theory or a low-energy limit of Ho\v{r}ava gravity,
it was argued of the radiations at the universal horizon, where even an {\it infinite-speed} particle can not escape
from \cite{Berg:2012}, but none at
the Killing horizon in \cite{Berg:2012PRL}.
On the contrary, in \cite{Mich:2015},
it was shown the opposite results in a more direct calculation, {\it i.e.},
the radiations at the Killing horizon but none at the universal horizon. This has
been a long-standing puzzle and it has been  clarified only recently that the
disagreements were due to the different choices observer's frames (or vacuum)
in \cite{Herr:2020}.

The result of \cite{Mich:2015}
seems to support the analysis on black hole thermodynamics in this paper since it shows a thermal radiation ``at infinity" with a Hawking temperature fixed by the Killing horizon, not the universal horizon, as suggested in (15).
Actually, for {\it relativistic} particles, the Killing horizon has the same role as in GR and produces the Hawking radiations as usual: For  $(3+1)$-dimensional non-rotating black holes in the (UV-complete) Ho\v{r}ava gravity,
see \cite{Peng:2009}.
Moreover, interestingly, it is found that the new horizon at
$r=0$ corresponds to a universal horizon
\cite{Soti:2014}
and its vanishing temperature for flat or dS case in (\ref{T_C_zero}) supports
no Hawking radiations at the universal horizons in \cite{Mich:2015}. However,
its full understanding, especially on the (modified) black hole thermodynamics
at the Killing horizon, seems to be still an open problem.}

\section*{Acknowledgments}

I would like to thank Deniz O. Devecioglu, Kyung Kiu Kim, and Cheong Oh Lee for helpful discussions and numerical plotting of Fig. 6. This was supported by Basic Science Research Program through the National Research Foundation of Korea (NRF) funded by the Ministry of Education, Science and Technology {(2020R1A2C1010372, 2020R1A6A1A03047877)}.

\newcommand{\J}[4]{#1 {\bf #2} #3 (#4)}
\newcommand{\andJ}[3]{{\bf #1} (#2) #3}
\newcommand{\AP}{Ann. Phys. (N.Y.)}
\newcommand{\MPL}{Mod. Phys. Lett.}
\newcommand{\NP}{Nucl. Phys.}
\newcommand{\PL}{Phys. Lett.}
\newcommand{\PR}{Phys. Rev. D}
\newcommand{\PRL}{Phys. Rev. Lett.}
\newcommand{\PTP}{Prog. Theor. Phys.}
\newcommand{\hep}[1]{ hep-th/{#1}}
\newcommand{\hepp}[1]{ hep-ph/{#1}}
\newcommand{\hepg}[1]{ gr-qc/{#1}}
\newcommand{\bi}{ \bibitem}


\begin{thebibliography}{999}

\bibitem{Park:1207}
  M.~I.~Park,
  Phys.\ Lett.\ B {\bf 718}, 1137 (2013) [erratum: Phys. Lett. B \textbf{809}, 135720 (2020)]
  [arXiv:1207.4073 [hep-th]].

\bi{Soti}
  T.~P.~Sotiriou, M.~Visser and S.~Weinfurtner,
 Phys.\ Rev.\ D {\bf 83}, 124021 (2011)
  [arXiv:1103.3013 [hep-th]].

\bibitem{Lifs} E. M. Lifshitz, Zh. Eksp. Teor. Fiz., {\bf 11}, 255 $\&$ 269 (1941).

\bibitem{DeWi}
  B.~S.~DeWitt,
  Phys.\ Rev.\  {\bf 160}, 1113 (1967).

\bibitem{Hora}
  P.~Ho\v{r}ava,
  JHEP {\bf 0903}, 020 (2009) [arXiv:0812.4287 [hep-th]];
  Phys.\ Rev.\  D {\bf 79}, 084008 (2009)
  [arXiv:0901.3775 [hep-th]].

\bibitem{Bana}
  M.~Banados, C.~Teitelboim and J.~Zanelli,
  Phys.\ Rev.\ Lett.\  {\bf 69}, 1849 (1992)
    [hep-th/9204099].

\bibitem{Park:9806}
  M.~I.~Park,
    Phys.\ Lett.\ B {\bf 440}, 275 (1998)
      [hep-th/9806119].

\bibitem{Kerr} R. P. Kerr, Phys. Rev. Lett. {\bf 11}, 237 (1963).

\bibitem{Penr} R. Penrose, Riv. Nuovo Cim. {\bf 1} 252 (1969).

\bibitem{Dese:1982} S. Deser, R. Jackiw, and S. Templeton,
Phys.\ Rev.\ Lett.\  {\bf 48}, 975 (1982).

\bibitem{Berg:2009}
  E.~A.~Bergshoeff, O.~Hohm and P.~K.~Townsend,
  Phys.\ Rev.\ Lett.\  {\bf 102}, 201301 (2009)
  [arXiv:0901.1766 [hep-th]].

\bibitem{Dese}
  S.~Deser, R.~Jackiw and G.~'t Hooft,
  Annals Phys.\  {\bf 152}, 220 (1984).

\bibitem{Mann}
  R.~B.~Mann, J.~J.~Oh and M.~I.~Park,
  Phys.\ Rev.\ D {\bf 79}, 064005 (2009)
  [arXiv:0812.2297 [hep-th]].






\bi{Kiri}
  E.~B.~Kiritsis and G.~Kofinas,
  JHEP {\bf 1001}, 122 (2010)
  [arXiv:0910.5487 [hep-th]];
  G.~Koutsoumbas and P.~Pasipoularides,
  Phys.\ Rev.\ D {\bf 82}, 044046 (2010) [arXiv:1006.3199 [hep-th]].

\bibitem{Blas:2009}
  D.~Blas, O.~Pujolas and S.~Sibiryakov,
  Phys.\ Rev.\ Lett.\  {\bf 104}, 181302 (2010)
  [arXiv:0909.3525 [hep-th]].

\bibitem{Donn:2011}
  W.~Donnelly and T.~Jacobson,
  Phys.\ Rev.\ D {\bf 84}, 104019 (2011)
  [arXiv:1106.2131 [hep-th]];
  J.~Bellorin and A.~Restuccia,
  Phys.\ Rev.\ D {\bf 84}, 104037 (2011)
  [arXiv:1106.5766 [hep-th]];
  X.~Gao,
  Phys.\ Rev.\ D {\bf 90}, 104033 (2014)
  [arXiv:1409.6708 [gr-qc]].

\bibitem{Deve:2020}
  D.~O.~Devecioglu and M.~I.~Park,
 Eur. Phys. J. C {\bf 80}, 597 (2020) [arXiv:2001.02556 [hep-th]].

\bibitem{Kim:2007}
H.~C.~Kim, M.~I.~Park, C.~Rim and J.~H.~Yee,
JHEP \textbf{10}, 060 (2008)
[arXiv:0710.1362 [hep-th]].

\bibitem{Park:0602}
M.~I.~Park,
  Phys.\ Lett.\  B {\bf 647}, 472 (2007)
   [arXiv:hep-th/0602114];
  Phys.\ Rev.\  D {\bf 77}, 026011 (2008)
  [arXiv:hep-th/0608165];
  Phys.\ Rev.\  D {\bf 77}, 126012 (2008)
  [arXiv:hep-th/0609027];
  Phys.\ Lett.\  B {\bf 663}, 259 (2008)
   [arXiv:hep-th/0610140];
  Class.\ Quant.\ Grav.\  {\bf 25}, 095013 (2008)
 [arXiv:hep-th/0611048].


 \bibitem{Gibb}
  G.~W.~Gibbons and S.~W.~Hawking,
  Phys.\ Rev.\ D {\bf 15}, 2738 (1977).

\bibitem{Park:0705}
  M.~I.~Park,
  Class.\ Quant.\ Grav.\  {\bf 25}, 135003 (2008)
  [arXiv:0705.4381 [hep-th]].

\bibitem{Cvet:2018}
  M.~Cvetic, G.~W.~Gibbons, H.~Lu and C.~N.~Pope,
  Phys.\ Rev.\ D {\bf 98}, 
  106015 (2018)
  [arXiv:1806.11134 [hep-th]].




\bibitem{Corn}
  L.~Cornalba and M.~S.~Costa,
  Phys.\ Rev.\ D {\bf 66}, 066001 (2002)
  [hep-th/0203031];
  G.~Barnich, A.~Gomberoff and H.~A.~Gonzalez,
  Phys.\ Rev.\ D {\bf 86}, 024020 (2012)
  [arXiv:1204.3288 [gr-qc]].

{
\bibitem{Soti:2014}
T.~P.~Sotiriou, I.~Vega and D.~Vernieri,
Phys. Rev. D \textbf{90}, 
044046 (2014)
[arXiv:1405.3715 [gr-qc]].
}

\bibitem{Carl:1998}
  S.~Carlip,
  {\it Quantum gravity in 2+1 dimensions} (Cambridge Univ. Press, 1998)

\bibitem{Bell:2019}
  J.~Bellorin and B.~Droguett,
  Phys.\ Rev.\ D {\bf 100}, 
  064021 (2019)
  [arXiv:1905.02836 [gr-qc]].

\bibitem{Isra}
  W.~Israel,
  Phys.\ Lett.\ A {\bf 57}, 107 (1976);
  J.~M.~Maldacena,
  JHEP {\bf 0304}, 021 (2003)
  [hep-th/0106112];
  J.~Maldacena and L.~Susskind,
  Fortsch.\ Phys.\  {\bf 61}, 781 (2013)
  [arXiv:1306.0533 [hep-th]].

\bibitem{Stro}
  A.~Strominger,
  JHEP {\bf 0110}, 034 (2001)
  [hep-th/0106113];
  D.~Klemm,
  Nucl.\ Phys.\ B {\bf 625}, 295 (2002)
  [hep-th/0106247];
  V.~Balasubramanian, J.~de Boer and D.~Minic,
  Phys.\ Rev.\ D {\bf 65}, 123508 (2002)
  [hep-th/0110108].

\bibitem{Brow}
  J.~D.~Brown and J.~W.~York, Jr.,
  Phys.\ Rev.\ D {\bf 47}, 1407 (1993)
  [gr-qc/9209012];
  V.~Iyer and R.~M.~Wald,
  Phys.\ Rev.\ D {\bf 50}, 846 (1994)
  [gr-qc/9403028].

\bibitem{Argu}
  C.~Arguelles, N.~Grandi and M.~I.~Park,
  JHEP {\bf 1510}, 100 (2015)
  [arXiv:1508.04380 [hep-th]].

\bibitem{deBu}
  S.~de Buyl, S.~Detournay, G.~Giribet and G.~S.~Ng,
  JHEP {\bf 1402}, 020 (2014)
  [arXiv:1308.5569 [hep-th]].

\bibitem{Park:0801}
  M.~I.~Park,
  Phys.\ Rev.\ D {\bf 80}, 084026 (2009)
  [arXiv:0811.2685 [hep-th]].




\bibitem{Berg:2012}
P.~Berglund, J.~Bhattacharyya and D.~Mattingly,
Phys. Rev. D \textbf{85}, 124019 (2012)
[arXiv:1202.4497 [hep-th]];
J.~Bhattacharyya, M.~Colombo and T.~P.~Sotiriou,
Class. Quant. Grav. \textbf{33}, no.23, 235003 (2016)
[arXiv:1509.01558 [gr-qc]].

\bibitem{Berg:2012PRL}
P.~Berglund, J.~Bhattacharyya and D.~Mattingly,
Phys. Rev. Lett. \textbf{110},
071301 (2013)
[arXiv:1210.4940 [hep-th]];
B.~Cropp, S.~Liberati, A.~Mohd and M.~Visser,
Phys. Rev. D \textbf{89}, 
064061 (2014)
[arXiv:1312.0405 [gr-qc]].

\bibitem{Mich:2015}
F.~Michel and R.~Parentani,
Phys. Rev. D \textbf{91}, 
124049 (2015)
[arXiv:1505.00332 [gr-qc]].

\bibitem{Herr:2020}
M.~Herrero-Valea, S.~Liberati and R.~Santos-Garcia,
JHEP \textbf{04}, 255 (2021)
[arXiv:2101.00028 [gr-qc]].

\bibitem{Peng:2009}
J.~J.~Peng and S.~Q.~Wu,
Eur. Phys. J. C \textbf{66}, 325 
(2010)
[arXiv:0906.5121 [hep-th]];
M.~Liu, J.~Lu and J.~Lu,
Class. Quant. Grav. \textbf{28}, 125024 (2011)
[arXiv:1108.0758 [hep-th]];
M.~Liu, L.~Liu, J.~Zhang, J.~Lu and J.~Lu,
Gen. Rel. Grav. \textbf{44}, 3139 
(2012).

\end{thebibliography}
\end{document}